# Initial and Boundary Conditions
# for the Lattice Boltzmann Method

P. A. Skordos

*Center for Nonlinear Studies*

*Los Alamos National Lab, B258, Los Alamos, NM 87545*

*and*

*Massachusetts Institute of Technology*

*545 Technology Square, NE43-432, Cambridge, MA 02139*

June 19, 1993

## Abstract

A new approach of implementing initial and boundary conditions for the lattice Boltzmann method is presented. The new approach is based on an extended collision operator that uses the gradients of the fluid velocity. The numerical performance of the lattice Boltzmann method is tested on several problems with exact solutions and is also compared to an explicit finite difference projection method. The discretization error of the lattice Boltzmann method decreases quadratically with finer resolution both in space and in time. The roundoff error of the lattice Boltzmann method creates problems unless double precision arithmetic is used.

# 1  Introduction

The lattice Boltzmann method is a numerical scheme for simulating viscous fluids that is motivated by kinetic theory [1, 2, 3, 4, 5, 6, 7]. It can be viewed either as a discretization of a simplified Boltzmann equation using a symmetric lattice [8] or as a Lagrangian finite difference scheme for the Navier Stokes equations [9]. The kinetic theory point of view is emphasized here. The lattice Boltzmann method represents the state of the fluid at a computational node using *a set of real numbers* which are called *populations* and are analogous to the microscopic density function of the Boltzmann equation. The populations are convected from one lattice site to another in discrete time steps, and are relaxed towards local equilibrium between every convection. The relaxation step or collision operator conserves mass and momentum (and energy for thermal models) just like a particle collision in kinetic theory.

The mapping from the populations $F_i$ to the fluid variables $\rho, V_x, V_y$ is a straightforward summation (equation 3 of section 2) and is performed in every step of the computation (see section 2). By contrast, the inverse mapping from the fluid variables $\rho, V_x, V_y$ to the populations $F_i$ is not simple to compute and has not been used for computational purposes until the present work. In this paper we use the inverse mapping in order to implement accurate initial and boundary conditions. The inverse mapping can be obtained theoretically by the Chapman-Enskog expansion, but in practice the truncated Chapman-Enskog expansion does not perform very well (section 7). In this paper we present a new method for calculating the inverse mapping that is both accurate and easy to use, and we demonstrate that our method provides accurate initial and boundary conditions.

Traditionally the inverse mapping from the fluid variables $\rho, V_x, V_y$ to the populations $F_i$ has been avoided in practice by using special techniques. In the case of initial conditions, for example when the fluid density and velocity $\rho, V_x, V_y$ are specified at time zero and the goal is to calculate $\rho, V_x, V_y$ at later times, the populations $F_i$ can be initialized equal to the equilibrium values $F_i^{\mathrm{eq}}$ which are known in terms of $\rho, V_x, V_y$. The error that results from this approximation can be overcome by discarding the first few steps and measuring the parameters of the flow afterwards (recalibrating the solution). This is often done in the literature without further discussion. A problem with this approach however is that it ends up solving a slightly different problem than the original initial value problem. By contrast, traditional methods such as finite differences do not need any recalibration. Thus, to put the lattice Boltzmann method on equal footing with other methods it is desirable to have an accurate means of calculating the populations $F_i$ from the initial values of $\rho, V_x, V_y$.

In the case of boundary conditions there are techniques that avoid the inverse mapping as in the case of initial conditions. In particular for solid wall boundaries the velocity of the fluid can be forced to zero by imposing a no-slip bounce-back of the populations $F_i$ (see [10] for a discussion of the actual location of the wall as a function of simulation parameters). In the case of boundary conditions with non-zero velocity, such as the driven cavity problem [11, page 199], the velocity at the boundary can be "controlled" by inserting momentum (forcing) in every step. Ad-hoc forcing is not very accurate however and it also requires recalibration of the simulation parameters. In the case of an arbitrary velocity specification at the boundary, such as the fluid flows of section 8, the forcing techniques and the recalibration become very difficult. Thus it is desirable to have an accurate means of calculating the populations $F_i$ at a boundary node from the fluid variables $\rho, V_x, V_y$ that are specified at this node.

In this paper we present a new method for calculating the populations $F_i$ at any node from the fluid variables $\rho, V_x, V_y$ that are specified at this node. We have tested our method in the case of initialization and in the case of boundary conditions with arbitrary velocity and with zero velocity, and we have obtained good results in all cases.

For exposition we use the hexagonal 7-speed model of the lattice Boltzmann method with the $1/\tau$ collision operator (references [1, 2, 3, 4]), but it should be clear that our results apply to all lattice Boltzmann methods. For example it is straightforward to apply our results to the orthogonal 9-speed model, and this is done in the appendix. We present computer simulations using both the hexagonal 7-speed model and the orthogonal 9-speed model. In the next two sections we review the hexagonal model using notation that is suitable for our approach. Then we compare different ways of calculating the populations $F_i$ from $\rho, V_x, V_y$ including our new method. In section 6 we discuss the numerical roundoff error of the lattice Boltzmann method. We show that the numerical roundoff in the equilibrium population formulas causes problems unless double precision arithmetic is used. In sections 7 and 8 we discuss our simulations of initial value problems and boundary value problems. We compare the lattice Boltzmann method against an explicit finite difference projection method, and we also demonstrate that the lattice Boltzmann method has second order convergence both in space and in time.

## 2    Hexagonal 7-speed $1/\tau$ model

We consider a hexagonal lattice with six moving populations denoted by $F_i$ $i = 1, \ldots, 6$ and one rest-particle population denoted by $F_0$. We suppose that the fluid variables $\rho, V_x, V_y$ are known at every node at time zero, and that the goal is to calculate $\rho, V_x, V_y$ at later times. At startup the populations $F_i$ must be initialized from the given fluid variables $\rho, V_x, V_y$ (see section 4). After initialization, successive steps of relaxation and convection are performed to calculate the fluid variables $\rho, V_x, V_y$ at later times. The relaxation and convection steps are described by the following formulas,

$$
\begin{aligned}
F_i(\vec{x} + \vec{e_i}\,\Delta t,\, t + \Delta t) &= F_i(\vec{x}, t) + (-1/\tau)\,[F_i(\vec{x}, t) - F_i{}^{\text{eq}}(\vec{x}, t)] \\
F_0(\vec{x},\, t + \Delta t) &= F_0(\vec{x}, t) + (-1/\tau)\,[F_0(\vec{x}, t) - F_0{}^{\text{eq}}(\vec{x}, t)] \\
i &= 1, \ldots, 6
\end{aligned}
\tag{1}
$$

$$
\tau = \frac{1}{2} + \frac{4\Delta t\,\nu}{\Delta x^2} \quad .
$$

The relaxation parameter $\tau$ is chosen to achieve the desired kinematic viscosity $\nu$ given the space and time discretization parameters $\Delta x, \Delta t$. The vector $\vec{e_i}$ stands for the six velocity directions of the hexagonal lattice,

$$
\vec{e_i} = \frac{\Delta x}{\Delta t}\left(\cos\frac{2\pi(i-1)}{6},\ \sin\frac{2\pi(i-1)}{6}\right) \quad .
\tag{2}
$$

The velocity $\vec{V}(\vec{x}, t)$ and density $\rho(\vec{x}, t)$ are computed from the populations $F_i(\vec{x}, t)$ using the relations,

$$
\begin{aligned}
\rho(\vec{x}, t) &= \sum_{i=0}^{6} F_i(\vec{x}, t) \\
\rho(\vec{x}, t)\,\vec{V}(\vec{x}, t) &= \sum_{i=1}^{6} F_i(\vec{x}, t)\,\vec{e_i}
\end{aligned}
\tag{3}
$$

The variations of density around its mean value (spatial mean which is constant in time) provide an estimate of the fluid pressure $P(\vec{x}, t)$, according to the following equation,

$$
P(\vec{x}, t) = c_s^2\,(\rho(\vec{x}, t) - <\rho>) \quad .
\tag{4}
$$

The speed of sound is,

$$
c_s = \sqrt{3\,w_0}\,(\Delta x/\Delta t)
\tag{5}
$$

where the coefficient $w_0$ is discussed below. The equilibrium populations $F_i^{\mathbf{eq}}(x, t)$ are given by the following equations,

$$
\begin{aligned}
F_i^{\mathbf{eq}}(\vec{x}, t) &= \rho(\vec{x}, t)\ \left[ w_0 + w_1(\vec{e}_i \cdot \vec{V}) + w_{20}(\vec{e}_i \cdot \vec{V})(\vec{e}_i \cdot \vec{V}) + w_{21}(\vec{V} \cdot \vec{V}) \right] \\
F_0^{\mathbf{eq}}(\vec{x}, t) &= \rho(\vec{x}, t)\ \left[ z_0 + z_{21}(\vec{V} \cdot \vec{V}) \right]
\end{aligned}
\tag{6}
$$

$$
6\, w_0 + z_0 = 1 \ ,
$$
$$
w_1 = 1/(3\, c^2) \ , \quad w_{20} = 2/(3\, c^4) \ , \quad w_{21} = -1/(6\, c^2)
$$
$$
z_{21} = -1/c^2 \ , \quad c = \Delta x/\Delta t
$$

The above coefficients are chosen so that the Chapman-Enskog expansion (see section 3) of the evolution equation 1 matches the Navier Stokes equations. In particular the coefficient $w_1$ is determined from momentum conservation, the coefficient $w_{20}$ is determined from Galilean invariance (ie. the convection term $(V_x\, \partial V_x/\partial x + V_y\, \partial V_x/\partial y)$ must appear in the Chapman-Enskog expansion with a constant factor equal to one), the coefficient $w_{21}$ is chosen to eliminate the $(\vec{V} \cdot \vec{V})$ dependence of the pressure, and the coefficient $z_{21}$ is chosen to eliminate the $(\vec{V} \cdot \vec{V})$ term in the mass conservation equation. There is some freedom in choosing the remaining coefficients $w_0$ and $z_0$, but they must satisfy $6\, w_0 + z_0 = 1$ to conserve mass. In our simulations we use the balanced choice $w_0 = z_0 = (1/7)$ unless indicated otherwise.

The computation of the lattice Boltzmann method is organized as follows: The current lattice populations $F_i(\vec{x}, t)$ are used to calculate the velocity field $\vec{V}(\vec{x}, t)$ and density field $\rho(\vec{x}, t)$. These fields are the numerical solution at time $t$, and they are also used to compute the equilibrium populations $F_i^{\mathbf{eq}}(\vec{x}, t)$ which are needed to advance the solution. The equilibrium populations $F_i^{\mathbf{eq}}(\vec{x}, t)$ are used to relax the $F_i(\vec{x}, t)$ into "relaxed" populations which are then convected according to equation 1 to produce the lattice populations at the next time step. Then the cycle repeats. An implementation issue regarding roundoff error is discussed in section 6. Boundary conditions are discussed in section 4 and 8.

# 3 Chapman-Enskog Expansion

The Chapman-Enskog expansion is outlined here for completeness. The goal of the Chapman-Enskog expansion is to derive a set of partial differential equations in terms of $\rho$ and $\rho\vec{V}$ that approximate the behavior of the lattice Boltzmann fluid in the limit of $\Delta x, \Delta t$ going to zero, with the speed ratio $\Delta x/\Delta t = c$ constant. In particular we want to derive the mass continuity equation and the Navier Stokes momentum equations. The first step is to Taylor-expand the population variable $F_i(\vec{x} + \vec{e}_i\Delta t, t + \Delta t)$ in the evolution equation 1 around the point $(\vec{x}, t)$. This produces an equation whose left hand side is a Taylor series and whose right hand side is equal to $(-1/\tau)(F_i - F_i^{\mathrm{eq}})$. This equation has the following form to first order,

$$\Delta t \,(\partial F_i/\partial t + \vec{e}_i \cdot \nabla F_i) \,+\, \ldots = (-1/\tau)(F_i - F_i^{\mathrm{eq}}) \tag{7}$$

The second step is to combine the Taylor series equation 7 with the mass and momentum conservation relations (equation 3) to derive three equations corresponding to mass and momentum conservation. The left hand side of these equations is a Taylor series, and the right hand side is zero because the equilibrium populations $F_i^{\mathrm{eq}}$ are chosen to satisfy mass and momentum conservation (for example $\sum_0^6 F_i = \sum_0^6 F_i^{\mathrm{eq}}$). The three Taylor series that are derived in this way contain partial derivatives of quantities that are sums and tensors of the populations $F_i$. These equations have the following form to first order,

$$\begin{aligned}
\partial(\textstyle\sum_0^6 F_i)/\partial t \,+\, \vec{\nabla} \cdot (\textstyle\sum_1^6 \vec{e}_i F_i) \,+\, \ldots &= 0 \\
\partial(\textstyle\sum_1^6 \vec{e}_i F_i)/\partial t \,+\, \vec{\nabla} \cdot (\textstyle\sum_1^6 \vec{e}_i\vec{e}_i F_i) \,+\, \ldots &= 0
\end{aligned} \tag{8}$$

If we truncate the mass equation to first order terms in the derivatives, the resulting equation contains only sums of $F_i$ and no tensors. The sums of $F_i$ can be converted easily to $\rho$ and $\rho\vec{V}$, and this produces the mass continuity equation. The momentum equation must be truncated to second order terms in the derivatives to derive the Navier Stokes equations. This is necessary because second order spatial derivatives contribute to the viscosity of the fluid. Second order terms are not shown in equation 8 but they are easy to derive [12].

A complication arises with the pressure tensor $\left(\sum \vec{e_i}\vec{e_i}F_i\right)$ which appears in the momentum equation 8. The pressure tensor can not be expressed in terms of $\rho$ and $\rho\vec{V}$ without introducing an approximation of the $F_i$ in terms of $\rho$ and $\rho\vec{V}$. This approximation is necessary in the mass equation also if we include higher order terms in the mass equation. The approximation of the populations $F_i$ is the third step of the Chapman-Enskog expansion.

The Chapman-Enskog expansion approximates the populations $F_i(\vec{x}, t)$ with the equilibrium populations $F_i^{\mathrm{eq}}(\vec{x}, t)$ to zero order, and with the sum $F_i^{\mathrm{eq}}(\vec{x}, t) + F_i^{(1)}(\vec{x}, t)$ to first order. The correction term $F_i^{(1)}(\vec{x}, t)$ is discussed below. The approximation of the $F_i$ can be viewed as another series expansion that is used in parallel with the Taylor series expansion. Fortunately there is no need to calculate higher order approximations of $F_i$ than the first order approximation $F_i^{\mathrm{eq}}(\vec{x}, t) + F_i^{(1)}(\vec{x}, t)$ in order to retrieve the Navier Stokes equations. However we must calculate to second order terms in the Taylor series, as stated previously, in order to retrieve all the viscosity terms.

The correction term $F_i^{(1)}$ is computed from $F_i^{\mathrm{eq}}$ using the evolution equation 1 Taylor-expanded to first order with the $F_i$ replaced by the zero order estimate $F_i^{\mathrm{eq}}$ as follows,

$$F_i^{(1)} = -\tau\,\Delta t\left[\frac{\partial F_i^{\mathrm{eq}}}{\partial t} + \vec{e_i}\cdot\vec{\nabla}F_i^{\mathrm{eq}}\right] \tag{9}$$

Now we can use the estimate $F_i^{\mathrm{eq}}(\vec{x}, t) + F_i^{(1)}(\vec{x}, t)$ for the populations $F_i$ in the momen-tum equation 8, and we can also express $F_i^{eq}$ in terms of $\rho$ and $\rho\vec{V}$ to derive two partial differential equations in terms of $\rho$ and $\rho\vec{V}$ corresponding to momentum conservation. By choosing the parameters of the equilibrium population formulas appropriately, we can make the momentum equations match the Navier Stokes equations. For example the parameters of equation 6 produce the following x-momentum equation (to second order terms),

$$\frac{\partial(\rho V_x)}{\partial t} + \frac{\partial(\rho V_x V_x)}{\partial x} + \frac{\partial(\rho V_x V_y)}{\partial y} = -\frac{\partial(3c^2 w_0 \rho)}{\partial x} + \nu\,\nabla^2(\rho V_x) + \mu\,\frac{\partial(\vec{\nabla}\cdot(\rho\vec{V}))}{\partial x} \qquad (10)$$

$$\nu = \frac{c^2\,\Delta t}{8}\,(2\tau - 1) \qquad \mu = 2\,z_0\,\nu \quad .$$

Higher order terms (higher order derivatives of velocity and density) are discussed at some length in references [12, 13]. This concludes our summary of the Chapman-Enskog procedure.

# 4   Calculating the $F_i$ from $\rho, V_x, V_y$

We now address the problem of calculating the populations $F_i$ at any node from given values of the fluid variables $\rho, V_x, V_y$ at that node.

First we consider the **initial value problem** where the fluid variables $\rho, V_x, V_y$ are specified on every node at time zero, and the goal is to calculate $\rho, V_x, V_y$ at later times. If $\rho$ is not specified according to compressible fluid flow equations — for example $\rho$ is typically assumed to be constant in incompressible fluid flow — then $\rho$ must be computed from the pressure as follows,

$$\rho \;=\; <\rho> + (\frac{1}{c_s^2})\,P \qquad (11)$$

where $c_s$ is the speed of sound, $P$ is the pressure, and $<\rho>$ is the constant average density. It is very important **not** to initialize the density to be constant [14]. The density must follow the pressure gradients of the lattice Boltzmann fluid according to equation 11; otherwise large density waves and error transients will result.

Once the fluid variables $\rho, V_x, V_y$ are specified correctly, there are several ways of calculating an approximation to the populations $F_i$ using the fluid variables $\rho, V_x, V_y$. In this section we describe some of these approaches, including our new method, and in sections 7 and 8 we compare the different methods experimentally. For ease of reference we use keyword names inspired by [2] to refer to each method.

The first method, denoted by d2q7F0, approximates the population $F_i$ with the equilibrium value $F_i^{\text{eq}}$. As explained in section 1, this approximation is used very often in the literature, and it is accompanied by recalibration of the solution after the first few steps are discarded (initial transients). In our simulations we do not perform any recalibration however because our goal is to compare the accuracy of calculating the populations $F_i$ from the fluid variables $\rho, V_x, V_y$.

The second method, denoted by d2q7F1, is to use the first order approximation of the population $F_i$ given by the Chapman-Enskog expansion,

$$F_i = F_i^{\text{eq}} + F_i^{(1)}$$

$$F_i^{(1)} = -\tau \, \Delta t \left[ \frac{\partial F_i^{\text{eq}}}{\partial t} + \vec{e_i} \cdot \vec{\nabla} F_i^{\text{eq}} \right] \tag{12}$$

By differentiating the equilibrium population formulas (equation 6), we can get formulas for the derivatives of $F_i^{\text{eq}}$ in terms of the derivatives of the fluid variables $\rho, V_x, V_y$. The derivatives of $\rho, V_x, V_y$ may be known in some cases (for example in our exactly solvable fluid flow problems), but in general they must be estimated using finite differences. In our initialization tests of section 7 we use finite differences. In particular the time derivatives of $\rho, V_x, V_y$ are estimated using the Navier Stokes momentum and continuity equations, and the spatial derivatives of $\rho, V_x, V_y$ are estimated using spatial finite differences. We have also tested the initialization methods of this section using the exact values for the derivatives, and the results are qualitatively the same as those reported in section 7.

The third method for calculating the populations $F_i$ from the fluid variables $\rho, V_x, V_y$ is denoted by d2q7X and it is based on an **extended collision operator** that is presented in the next section. The extended collision operator controls the viscosity independent of the relaxation parameter $\tau$, so that $\tau$ can be set equal to one. **This means that the extended collision operator replaces the $F_i$ with $F_i^{\mathrm{eq}}$ in each step, and thus the $F_i$ can be set exactly equal to $F_i^{\mathrm{eq}}$ at startup and at the boundary nodes.**

The new method d2q7X is very accurate for implementing initial and boundary conditions as we shall see in section 7. However it is not very accurate when iterated many times. The reason is that the extended collision operator controls the viscosity using the gradients of the fluid velocity, and the gradients are computed using finite differences. The inexactness of finite differences produces an error in viscosity which means that the computed solution decays at a slightly different rate than desired. Thus, the error accumulates with successive iterations, and the method fails to converge as $\Delta t$ goes to zero (see figure 3 in section 7.2).

The best of both collision operators can be achieved however (standard method d2q7 and extended method d2q7X) by combining the two operators. We observe that two different collision operators having the same transport coefficients (shear and bulk viscosity in our case) can be used interchangeably in the same computation. We have verified experimentally that switching between the different but equivalent collision operators incurs only a small error that is negligible compared to other errors. Thus we propose a **new hybrid method** where the extended collision operator is used at the boundary nodes all the time, and at every node during the first step of the computation. After the first step, the standard collision operator is used at the inner (non-boundary) nodes. The new hybrid method is denoted by d2q7H, and it is our recommended method for the hexagonal 7-speed model. The implementation of boundary conditions using the hybrid method is discussed further in section 8.

# 5    Extended Collision Operator

The idea behind the new collision operator is to include additional terms in the equilibrium population formulas that are based on the gradients of the fluid velocity, so that the viscosity can be controlled independent of the relaxation parameter $\tau$. The terms that we use are motivated by equation 2.5.1 in reference [12].

The new collision operator is used in the same way as the standard collision operator; namely the evolution equation and the conservation relations (equations 1 and 3) remain unchanged. The new equilibrium population formulas are as follows,

$$
\begin{aligned}
F_i^{\text{eq}}(\vec{x}, t) &= \rho(\vec{x}, t) \left[ w_0 + w_1(\vec{e}_i \cdot \vec{V}) + w_{20}(\vec{e}_i \cdot \vec{V})(\vec{e}_i \cdot \vec{V}) + w_{21}(\vec{V} \cdot \vec{V}) \right] + \\
&\quad w_{31} \left( \vec{e}_i \cdot \vec{\nabla}(\vec{e}_i \cdot \rho\vec{V}) \right) + w_{32} \left( \vec{\nabla} \cdot \rho\vec{V} \right) \quad, \\
&\quad i = 1, \ldots, 6 \\
F_0^{\text{eq}}(\vec{x}, t) &= \rho(\vec{x}, t) \left[ z_0 + z_{21}(\vec{V} \cdot \vec{V}) \right] + z_{32} \left( \vec{\nabla} \cdot \rho\vec{V} \right) \\
& \qquad\qquad 3\, c^2\, w_{31} + 6\, w_{32} + z_{32} = 0
\end{aligned}
$$

(13)

The velocity gradients in the above equation (the terms with coefficients $w_{31}, w_{32}, z_{32}$) are computed using finite differences unless they are known by other means; for example the velocity gradients may be known at the boundary nodes (see section 8). The coefficients $w_0, w_1, w_{20}, w_{21}, z_0, z_{21}$ have the same values as in the standard collision operator d2q7 (equation 6). The shear and bulk viscosity of the extended collision operator are given by the following formulas (calculated using the Chapman-Enskog procedure),

$$\nu = \frac{c^2 \, \Delta t}{8} \left( 2\tau - 1 \right) \; - \; \frac{3 \, c^4 \, w_{31}}{4} \tag{14}$$

$$\mu = 2 \, z_0 \, \nu \; - \; \frac{3 \, c^4 \, w_{31}}{2} \left( 1 - z_0 \right) \; - \; 3c^2 w_{32}$$

Once the relaxation parameter $\tau$ is set equal to one, the coefficient $w_{31}$ is chosen to achieve the desired kinematic viscosity $\nu$ given the discretization parameters $\Delta x, \Delta t$. The coefficient $w_{32}$ is chosen to achieve the desired bulk viscosity. In the case of the hybrid method d2q7H discussed in section 4, the bulk viscosity of equation 14 is chosen equal to the bulk viscosity of the standard collision operator given by equation 10. The coefficient $z_{32}$ must satisfy $(3 \, c^2 \, w_{31} + 6 \, w_{32} + z_{32}) = 0$ so that the equilibrium populations conserve mass.

An extented collision operator for the orthogonal 9-speed model is easy to derive and is presented in the appendix.

# 6    Numerical Roundoff

This section discusses an implementation issue of the lattice Boltzmann method that can cause problems if one is unaware of it. If the lattice Boltzmann method is implemented exactly as described by equations (1,3,6), then the method suffers from roundoff error that grows as the ratio $(V/c)$ becomes small; that is as $(\Delta x / \Delta t)$ becomes large. This is undesirable because large values of $(\Delta x / \Delta t)$ are often used to improve the accuracy of the lattice Boltzmann method.

| term | $w_0$ | $w_1$ | $w_{20}$ | $w_{21}$ |
|------|-------|-------|----------|----------|
| size | 1 | $V/c$ | $V^2/c^2$ | $V^2/c^2$ |

Table 1: The terms of the equilibrium population formula have different sizes. When they are added together, numerical roundoff error can be significant.

The roundoff error (numerical loss of precision) arises in the computation of the equilibrium populations using equation 6. This formula is a sum of four terms. If we factor out the density $\rho(\vec{x}, t)$, the first term is a constant coefficient $w_0$ and the remaining terms are proportional to $V/c$, $V^2/c^2$, and $V^2/c^2$ respectively (see table 1). Consequently when $V/c$ is small, for example $V/c \simeq 10^{-3}$, the terms to be added have very disparate sizes and their sum suffers a significant loss of accuracy when the computer aligns the numbers to be added (about 5 or 6 decimal places when $V/c \simeq 10^{-3}$). If single precision arithmetic is used (about eight decimal places), then the loss of five digits is a serious problem.

In section 7.3 we will see that the computational error of the lattice Boltzmann method decreases at first when the speed $\Delta x / \Delta t$ increases, but after some point the error starts to increase with higher speeds. For example in the Taylor vortex when the maximum fluid speed is 1.0, the error starts to increase at the rate of $(\Delta x / \Delta t)^{1.4}$ when the microscopic speed ($\Delta x / \Delta t$) is larger than 300. Fortunately the error growth disappears when double precision arithmetic is used, and this confirms that the breakdown of the method is caused by numerical roundoff. Furthermore we can estimate that each additional decimal place of computer arithmetic delays the roundoff problem in the equilibrium populations by a factor of 10 in the speed $\Delta x / \Delta t$. **This means that double precision arithmetic eliminates** **the roundoff problem for most practical purposes.**

We also note that apart from using double precision arithmetic, there is an algebraic transformation that reduces the roundoff error in the equilibrium populations, and it can be used in all cases because it does not involve any additional cost. The algebraic transformation does not eliminate the roundoff error completely however; in general double precision arithmetic remains necessary. The idea is to modify the populations $F_i$ defined by equations 1, 3, 6 as follows,

$$\begin{aligned}
\widehat{F_i} &= F_i - w_0 <\rho> \\
\widehat{F_i^{\text{eq}}} &= F_i^{\text{eq}} - w_0 <\rho>
\end{aligned} \tag{15}$$

where the spatial average density $<\rho>$ is constant in time and typically equal to one. The non-moving population become $\widehat{F_0} = F_0 - z_0 <\rho>$. The conservation relations are modified accordingly,

$$\begin{aligned}
\rho(\vec{x}, t) &= \sum_{i=0}^{6} \widehat{F_i}(\vec{x}, t) + <\rho> \\
\rho(\vec{x}, t)\, \vec{V}(\vec{x}, t) &= \sum_{i=1}^{6} \vec{e_i}\, \widehat{F_i}(\vec{x}, t) \quad .
\end{aligned} \tag{16}$$

The new equilibrium population formulas are as follows,

$$\begin{aligned}
\widehat{F_i^{\text{eq}}}(\vec{x}, t) &= w_0\left(\rho(\vec{x}, t) - <\rho>\right) + \\
&\quad \rho(\vec{x}, t)\left[w_1(\vec{e_i}\cdot\vec{V}) + w_{20}(\vec{e_i}\cdot\vec{V})(\vec{e_i}\cdot\vec{V}) + w_{21}(\vec{V}\cdot\vec{V})\right]
\end{aligned} \tag{17}$$

$$\widehat{F_0^{\text{eq}}}(\vec{x}, t) = z_0\left(\rho(\vec{x}, t) - <\rho>\right) + \rho(\vec{x}, t)\, z_{21}\,(\vec{V}\cdot\vec{V})$$

The new equilibrium population formulas are numerically better than the original ones because the term that used to be $w_0\,\rho$ is now $w_0\,(\rho - <\rho>)$. The new quantity $(\rho - <\rho>)$ is of the order $P/(3c^2 w_0)$ and the pressure $P$ is of the order $\rho\, V^2$ as can be seen from the Navier Stokes equations. Hence the expression $w_0\,(\rho - <\rho>)$ is of the order $\rho(V^2/c^2)$. The new formulas compute the same quantities as the original formulas, and they incur a smaller loss of precision. Loss of precision still occurs when the terms proportional to $(V/c)$ and $(V^2/c^2)$ are combined.

# 7  Results — Initial Value

First we test **initial value problems**. For this purpose we use the analytic solutions of a decaying Taylor vortex and a decaying shear flow in two dimensions with periodic boundary conditions. Figure 1 shows the velocity vector fields of these flows.

The decaying Taylor vortex [15] has the following analytic solution,

$$
\begin{aligned}
V_x(x,y,t) &= (-1/A)\,\cos(Ax)\,\sin(By)\,\exp(-2\alpha\,\nu t) \\
V_y(x,y,t) &= (1/B)\,\sin(Ax)\,\cos(By)\,\exp(-2\alpha\,\nu t) \\
P(x,y,t) &= -(1/4)\,[\cos(2Ax)/A^2 \,+\, \cos(2By)/B^2]\,\exp(-4\alpha\,\nu t) \quad ,
\end{aligned}
\tag{18}
$$

where the constant $\alpha$ is equal to $(A^2 + B^2)/2$, and $\nu$ is the kinematic viscosity. The length constants $A, B$ are chosen $A = 1$ and $B = 2/\sqrt{3}$ to produce the **hexagonal Taylor vortex**, and $A = B = 1$ to produce the **orthogonal Taylor vortex**. The former is used to test the hexagonal 7-speed model, and the latter is used to test the orthogonal 9-speed model. The flow region of the hexagonal Taylor vortex is $0 <= x <= 2\pi$ and $0 <= y <= \pi\sqrt{3}$, and can be covered exactly by a hexagonal lattice using periodic boundary conditions. Similarly, the flow region of the orthogonal Taylor vortex is $0 <= x <= 2\pi$ and $0 <= y <= 2\pi$, and can be covered exactly by an orthogonal lattice using periodic boundary conditions.

The decaying shear flow has the following analytic solution,

$$
\begin{aligned}
V_x(x, y, t) &= A \\
V_y(x, y, t) &= B \, \cos(k \, x - k \, A \, t) \, \exp(-k^2 \, \nu \, t) \\
P(x, y, t) &= \text{constant} \quad,
\end{aligned}
\tag{19}
$$

where the constant $k$ is chosen $k = 1$ so that $x$ varies between $0 \mathrel{<=} x \mathrel{<=} 2\pi$, and the length constants $A, B$ are chosen $A = B = 1$ so that the horizontal velocity is equal to the maximum vertical velocity. The vertical extent of the shear flow is chosen $0 \mathrel{<=} y \mathrel{<=} \pi\sqrt{3}$ for the hexagonal case, and $0 \mathrel{<=} y \mathrel{<=} 2\pi$ for the orthogonal case in complete analogy with the Taylor vortex.

In all of the results below the coefficient of shear viscosity is chosen equal to one, $\nu = 1$. The measured error $V^E$ denotes the velocity relative error, and is calculated according to the following formula,

$$
V^E = \frac{\sum_{x,y} |V_x - V_x^*|}{\sum_{x,y} |V_x^*|} + \frac{\sum_{x,y} |V_y - V_y^*|}{\sum_{x,y} |V_y^*|}
\tag{20}
$$

where $V^*$ denotes the exact analytic solution, and the sums are taken over the whole grid. In the case of the Hagen-Poiseuille flow and the oscillating plate problem (see section 8) where $\sum_{x,y} |V_y^*| = 0$, we use a different normalization as follows,

$$
V^E = \frac{\sum_{x,y} |V_x - V_x^*| + \sum_{x,y} |V_y - V_y^*|}{\sum_{x,y} |V_x^*|}
\tag{21}
$$

Double precision arithmetic is used in all of the reported results unless stated otherwise (for example in figure 4).

We define the Mach number $M$ using the maximum fluid speed at time zero, which is equal to 1.0 for all of our test cases,

$$
M = 1/c_s = \Delta t / (\Delta x \sqrt{3 w_0})
\tag{22}
$$

We also define the pseudo-Mach number or "computational Mach number" $M_c$,

$$M_c = 1/c = \Delta t / \Delta x \qquad (23)$$

and we use $M_c$ rather than $M$ in our figures because the discretization error of the lattice Boltzmann method depends on $M_c$ rather than $M$ as we will see below. In the case of the Taylor vortex, which is a solution of the incompressible Navier Stokes equations, the choice $w_0 = 1/7$ produces a Mach number small enough that the compressible effects are smaller than the discretization error. In the case of shear flow, which has zero density gradient and is a solution of the compressible Navier Stokes equations, the error is independent of the Mach number $M$ and it depends only on $M_c$.

For the hexagonal 7-speed model the choice $w_0 = 1/7$ produces a Mach number that varies with the discretization parameters $\Delta x, \Delta t$ as follows $M = 1.53\, M_c = 1.53\Delta t/\Delta x$. For the orthogonal 9-speed model the choice $y_0 = w_0/4$ and $w_0 = 1/7$ (see appendix) gives $M = 1.53\, M_c$ also. Another choice $w_0 = 10^{-6}/3$ is discussed briefly in section 7.4 for the purpose of allowing high Mach numbers with small $M_c$ in particular $M = 10^3\, M_c$. Different values of $w_0$ are used in section 7.5 for the purpose of plotting the error of the lattice Boltzmann method as a function of $\Delta t$ while keeping the Mach number constant. The Mach number is kept constant by varying $w_0$ in proportion to $\Delta t^2$. This study allows us to distinguish between compressible effects and the intrinsic discretization error of the lattice Boltzmann method. Details can be found in section 7.5.

## 7.1   Initialization error

This section compares the different methods of initialization that are described in section 4 and are denoted by d2q7F0, d2q7F1, d2q7X, d2q7H. We note that d2q7X and d2q7H are identical for the first step because they both use the extended collision operator for initialization. After the first step d2q7H uses the standard collision operator at the inner (non-boundary) nodes, and d2q7X continues to use the extended collision operator at every node. Every node is an inner node in this experiment because periodic boundary conditions are used.

Figure 2 plots the error as it develops during the first 10 steps of the computation. A $30 \times 30$ grid is used ($\Delta x = 2\pi/30 = 0.2094$). Figure (a) plots the error in the case of the hexagonal Taylor vortex, using $\Delta t = 0.001$ which gives $\tau = 0.5912$ for the standard collision operator (equation 1). The curves shown correspond to d2q7F0, d2q7F1, d2q7X, d2q7H (solid, dashed, dotted, dash-dotted lines). Figure (b) plots the same data using $\Delta t = 0.025$ which gives $\tau = 2.780$ for the standard collision operator. We can see that the extended collision operator, which is employed by schemes d2q7X and d2q7H, provides the best results in all cases. We can also see that the first order correction scheme d2q7F1 is more accurate than the zero order correction scheme d2q7F0 when $\tau < 1$ and inversely when $\tau > 1$. This indicates that the first order correction given by equation 12 becomes inaccurate when $\tau$ is large. Figures (c) and (d) plot the same data as figures (a) and (b) but for the case of shear flow. The results are qualitatively the same.

These experiments demonstrate that the extended collision operator can be used to initialize accurately the populations $F_i$ from the fluid variables $\rho, V_x, V_y$ in an initial value problem.

## 7.2 Iterating the extended collision operator

We now examine the performance of the extended collision operator when iterated many times. We recall that the extended collision operator uses the gradients of the fluid velocity in order to control the viscosity. Figure 3 shows the error in simulating the hexagonal Taylor vortex and the hexagonal shear flow using a $30 \times 30$ grid. The error is plotted against $M_c$ with $\Delta t$ varying, and is calculated at the final time $T = 1.0$ when the maximum velocity of the hexagonal Taylor vortex is approximately $1/10$ of its initial value. The curves correspond to the hybrid method d2q7H, the extended collision operator d2q7X using finite differences to calculate the gradients, and again the extended collision operator d2q7X using the known analytic solution to calculate the gradients (solid, dashed, dotted lines). When the curves of figure 3 intersect at $M_c = 0.026$, the relaxation parameter $\tau$ of the standard collision operator is equal to one, and the coefficients $w_{31}, w_{32}, z_{32}$ of the extended collision operator vanish. At this point the extended collision operator is identical to the standard collision operator.

As $M_c$ decreases below the value $M_c = 0.026$, the error of extended collision operator d2q7X using finite differences to calculate the gradients begins to grow and reaches relative error one as $M_c$ goes to zero (dashed line). By contrast, the error of the extended collision operator d2q7X using the analytic solution to calculate the gradients decreases towards a minimum error (dotted line) which is determined by the spatial discretization error of the $30 \times 30$ grid. This shows that the use of finite differences creates problems after repeated iterations. As explained in section 4 the inexactness of finite differences produces an error in viscosity which accumulates and becomes large after repeated iterations.

The hybrid method d2q7H does not suffer from the problems of the extended collision operator after repeated iterations because the hybrid method uses the standard collision operator at the inner nodes after the first step. For example in the case of periodic boundary conditions all the nodes are inner, and the hybrid method uses the extended collision operator only for initialization. Figure 3 shows that the hybrid method performs very well in the case of periodic boundary conditions and remains accurate as $M_c$ goes to zero (solid line). On the other hand, in the case of an actual boundary the hybrid method uses the extended collision operator at the boundary throughout the computation. In section 8 we will see that the use of finite differences at the boundary does not cause any problems as $M_c$ goes to zero, but it may cause instabilities when $M_c$ is large. The instabilities can be avoided by specifying a Neumann condition for the velocity at the boundary nodes (see section 8).

## 7.3   Roundoff error

Figure 4 compares the error of the lattice Boltzmann method (d2q7H version) when single precision arithmetic is used, when single precision arithmetic together with the algebraic transformation of section 6 is used, and when double precision arithmetic is used (dotted, dashed, solid lines). The data comes from simulations of the hexagonal Taylor vortex with periodic boundary conditions and $30 \times 30$ grid. The error is plotted against $M_c$ with $\Delta t$ varying and is calculated at the final time $T = 1.0$. When single precision arithmetic is used and the speed $\Delta x/\Delta t$ exceeds 300 (therefore $M_c < 0.003$), there is a growth of error that is caused by numerical roundoff. The procedure of section 6 together with single precision arithmetic can reduce the roundoff error but it can not prevent it. Double precision arithmetic is necessary to prevent the error growth in the Taylor vortex for $M_c < 0.003$. As explained in section 6 double precision arithmetic eliminates the roundoff problem for most practical purposes.

## 7.4    Comparison with traditional method

We compare the error of the hybrid lattice Boltzmann method d2q7H and the error of an explicit finite difference projection method in simulating the hexagonal Taylor vortex and the hexagonal shear flow with periodic boundary conditions. Both of these flows are defined in the hexagonal region $0 <= x <= 2\pi$ and $0 <= y <= \pi\sqrt{3}$, which means that the finite difference projection method must use the discretization $\Delta y = \Delta x \sqrt{3}/2$. For ease of reference we denote the projection method by EP7 when it is applied to a hexagonal region, and by EP9 when it is applied to an orthogonal region (this is done in later sections).

The explicit finite difference projection method computes an estimate of the velocity by solving a discretized Navier Stokes momentum equation where the pressure term is omitted [11, page 160]. Then the velocity estimate is corrected in order to satisfy incompressibility by solving a Poisson equation. This correction takes into account the pressure effects that were omitted in the first estimate of the velocity. In addition the solution of the Poisson equation provides an estimate of the pressure at the current time-step. In our simulations we use SOR [16, page 680] to solve the Poisson equation, forward Euler to estimate the time derivative, and 3-point symmetric differences to calculate the spatial derivatives on a grid that is orthogonal and non-staggered.

Figure 5 (a) plots the error in simulating the hexagonal Taylor vortex against $M_c$ with $\Delta t$ varying. The error is calculated at the final time $T = 1.0$ when the maximum velocity of the hexagonal Taylor vortex is approximately $1/10$ of its initial value. The curves shown correspond to d2q7H using $30 \times 30$ grid, d2q7H using $60 \times 60$ grid, EP7 using $30 \times 30$ grid, and EP7 using $60 \times 60$ grid (solid, dashed, dotted, dash-dotted lines). Figure (b) plots the same data against the dimensionless parameter $\Delta t\, \nu / \Delta x^2$ which facilitates comparison between different grids. Figures (c) and (d) plot the same data for shear flow. We can see that the Taylor vortex triggers an instability in the explicit projection method EP7 when $\Delta t\, \nu / \Delta x^2 >= 0.2$, but the shear flow does not trigger an instability.

With regard to the lattice Boltzmann method we observe that it can not approximate the solution (has a relative error of 1.0) when $M_c$ is larger than 0.2 approximately. In the case of the Taylor vortex, which is a solution of the incompressible fluid flow equations, it may appear that the problem arises from the compressibility of the lattice Boltzmann fluid (when $M_c \sim 0.2$, the Mach number is approximately $M = 1.53\, M_c = 0.3$). In the case of the shear flow however compressibility is not important. The shear flow is a solution of the compressible fluid flow equations, and it should be easily computed by the lattice Boltzmann method both at low and high Mach numbers. In fact the shear flow can be simulated at high Mach numbers by using a smaller $w_0$, for example $w_0 = 10^{-6}/3$ (see below).

The limitations of the lattice Boltzmann method shown in figure 5 when $M_c$ is larger than 0.2 persist independent of the Mach number. The limitations stem from the fact that the microscopic speed of advection becomes comparable to the fluid speed when $M_c$ approaches 1.0 and the Chapman-Enskog expansion breaks down.

With regard to simulating shear flow at high Mach numbers, we can choose $w_0 = 10^{-6}/3$ which gives $M = 10^3\, M_c$. The error of the lattice Boltzmann method d2q7H in simulating shear flow with $M = 10^3\, M_c$ is identical to the error plotted in figure 5(c). The error in simulating shear flow is independent of the Mach number because the density gradients of the shear flow are zero.

## 7.5    Quadratic Convergence

This section shows that the lattice Boltzmann method has second order convergence both in space and in time. Second order convergence in space means that the error decreases quadratically with $\Delta x$ while keeping the dimensionless parameter $\Delta t \, \nu / \Delta x^2$ constant (see [17, page 75]). Second order convergence in time means that the error decreases quadratically with $\Delta t$ while keeping the space discretization $\Delta x$ constant. In addition we are interested in the intrinsic error that arises from finite discretization, and not the error that arises from compressibility. When using compressible fluid dynamics such as the lattice Boltzmann method to simulate incompressible flow such as the Taylor vortex, it is important to distinguish between the error that arises from compressibility and the error that arises from finite discretization. In figure 5 the Mach number decreases in proportion to $M_c$, and thus the effects of compressibility and finite discretization can not be distinguished without further investigation.

To distinguish between the effects of compressibility and discretization error, we perform the same simulations as those in figure 5, while keeping the Mach number constant and varying the density coefficient $w_0$ as follows,

$$w_0 \;=\; \frac{1}{3} \left( \frac{\Delta t}{\Delta x \, M} \right)^2 \qquad (24)$$

In figure 6 (a) we show the error of d2q7H in simulating the hexagonal Taylor vortex at constant Mach number $M = 0.02$ using a $30 \times 30$ grid and a $60 \times 60$ grid (two dashed lines). For comparison purposes we also show the error of d2q7H using constant $w_0 = 1/7$ and variable Mach number (two solid lines). The constant Mach number curves are identical to the constant $w_0$ curves except for instabilities which are discussed below. This indicates that the compressible effects at Mach number $M = 0.02$ are smaller than the discretization error of both the $30 \times 30$ and $60 \times 60$ grids. The instability of the constant Mach number curves (dashed lines) is expected and it occurs when the density coefficient $w_0$ given by equation 24 becomes greater than $1/6$ which forces the density coefficient $z_0$ to become negative. Similar instabilities can be seen in figure 6 (c) which plots the same data for shear flow at constant Mach number $M = 0.05$.

Figure 6 (b) shows the same data as figure 6 (a) while keeping the Mach number constant at $M = 0.1$. The discretization error of the $30 \times 30$ grid is once again larger than the compressible effects. However, the discretization error of the $60 \times 60$ grid is smaller than the compressible effects when $\Delta t \, \nu / \Delta x^2$ becomes less than 0.1 approximately. In other words, the minimum discretization error of the $60 \times 60$ grid can not be achieved when the Mach number is kept constant at $M = 0.1$ because the compressible effects are larger than the minimum discretization error of the $60 \times 60$ grid.

We conclude that the discretization error of the lattice Boltzmann method is larger than the compressible effects when the density parameter is chosen constant $w_0 = 1/7$ and the Mach number varies with the discretization parameters $\Delta x, \Delta t$ according to the relation $M = \Delta t / (\Delta x \sqrt{3 w_0})$. Furthermore, this means that the error curves of figure 5 correspond to the true discretization error of the lattice Boltzmann method and not the effects of compressibility. Thus we can examine figure 5 to determine how the discretization error of the lattice Boltzmann method decreases with finer resolution.

If we examine the logarithmic plots of figure 5, we see that the error decreases quadratically (slope $-2$) with $\Delta t$ until a minimum space discretization error is reached. In addition the error decreases by 4 when we go from the $30 \times 30$ grid to the $60 \times 60$ grid while keeping the dimensionless parameter $\Delta t \, \nu / \Delta x^2$ constant — see figures 5 (b) and 5 (d). In other words the lattice Boltzmann method has second order convergence both in time and in space. In section 8 we will verify the second order convergence in boundary value problems also.

The explicit finite difference projection method EP7 has first order convergence in time and second order convergence in space. The first order convergence in time of the projection method EP7 can be seen easily in figures 5 (c) and 5 (d) which plot the error in simulating shear flow.

## 7.6    7-speed versus 9-speed

We now compare the accuracy of the hexagonal 7-speed model against the accuracy of the orthogonal 9-speed model. Figure 7 shows the error of d2q7H applied to the hexagonal Taylor vortex, and the error of d2q9H applied to the orthogonal Taylor vortex (solid and dashed lines). In addition the error of the explicit finite difference projection method is shown when applied to the hexagonal Taylor vortex with $\Delta y = \Delta x \sqrt{3}/2$ and also the orthogonal Taylor vortex with $\Delta y = \Delta x$ (dotted and dash-dotted lines). A $30 \times 30$ grid is used, and the error is calculated at the final time $T = 1.0$. We can see that the explicit finite difference projection method performs similarly on the hexagonal and the orthogonal Taylor vortices. On the other hand, the orthogonal 9-speed model d2q9H is significantly more accurate than the hexagonal 7-speed model d2q7H. A simple explanation is that nine speeds per node provide a better discretization of the microscopic velocity [8] than seven speeds per node.

# 8 Results — Boundary Value

In this section the orthogonal 9-speed model d2q9H is applied to several boundary value problems with exact solutions, and is also compared against the explicit finite difference projection method EP9. The boundary value problems are the one-quarter Taylor vortex, the Hagen-Poiseuille flow, and the oscillating plate above a stationary wall. Figure 8 shows the velocity vector fields of these flows, and also indicates the boundary nodes of each flow by drawing a square around the boundary nodes. Figure 8 (c) is plotted at time $t = 0.4$ when the oscillating plate starts moving to the left while the fluid below is still moving to the right.

The one-quarter Taylor vortex is defined in the region $\pi/2 <= x <= 3\pi/2$ and $\pi/2 <= y <= 3\pi/2$. The exact solution is given by equation 18 with $A = B = 1$. The boundary conditions are computed by evaluating the exact solution at the boundary lines, namely the horizontal and vertical lines $\pi/2 <= x <= 3\pi/2$ and $\pi/2 <= y <= 3\pi/2$. Dirichlet boundary conditions are used for both the velocity and the density. The density at the boundary nodes is calculated from the pressure using equation 11.

The Hagen-Poiseuille flow is defined in the region $0 <= x <= 1$ and $0 <= y <= 1$. The analytic solution is as follows,

$$
\begin{aligned}
V_x(x,y,t) &= -(y^2 - y)\,\Delta P \,/\,(2\nu) \\
V_y(x,y,t) &= 0 \\
P(x,y,t) &= (0.5 - x)\,\Delta P
\end{aligned}
\tag{25}
$$

The pressure gradient $\Delta P$ is chosen $\Delta P = 8.0$ so that the maximum fluid speed is 1.0 when $y = 1/2$. Dirichlet boundary conditions are used for both the velocity and the density. The values at the boundary nodes are computed by evaluating the exact solution at $0 <= x <= 1$ and $0 <= y <= 1$.

The oscillating plate problem is defined in the region $0 <= x <= 1$ and $0 <= y <= 1$ with periodic boundary conditions in the horizontal direction $x = 0$ and $x = 1$, and Dirichlet boundary conditions for the velocity at the top and bottom plates,

$$
\begin{aligned}
y = 1 : \quad & V_x = cos(\omega\, t) \quad & V_y = 0 \\
y = 0 : \quad & V_x = 0 \quad & V_y = 0
\end{aligned}
\tag{26}
$$

The density at the top and bottom plates can be specified as a Dirichlet condition, or it can be calculated dynamically (see section 8.1). However the results are identical in either case because the oscillating plate problem has constant density everywhere. The frequency of oscillation $\omega$ is chosen $\omega = 20$ so that the oscillating plate executes 3.18 cycles of oscillation during the time interval $T = 1.0$ which is used for testing (this is an arbitrary choice). The analytic solution of the oscillating plate problem [18, page 88] is given by the following equations,

$$
\begin{aligned}
V_x(x,y,t) \;=\; & (\cosh A \sin A(-2 \cosh B \sin B \cos \omega t + 2 \cos B \sinh B \sin \omega t) \\
& - \cos A \sinh A(2 \cosh B \sin B \sin \omega t + 2 \cos B \sinh B \cos \omega t)) \\
& / (\cos 2B - \cosh 2B) \\
V_y(x,y,t) \;=\; & 0 \\
P(x,y,t) \;=\; & \text{constant}
\end{aligned}
\tag{27}
$$

where $A = y\sqrt{\omega/(2\nu)}$ and $B = \sqrt{\omega/(2\nu)}$, and $\nu$ is the kinematic viscosity.

In the case of steady flow such as the Hagen-Poiseuille flow, we initialize $\rho, V_x, V_y$ equal to the exact steady state solution. Then we iterate for 100 steps, and test whether the fluid is in steady state. If the fluid is in steady state, we measure the velocity relative error $V^E$. Otherwise we keep iterating until the fluid reaches steady state. The goal of this procedure is to measure the error at steady state and not to characterize how quickly the fluid reaches steady state. Our criterion for steady state is that the relative change in velocity between successive iterations divided by $\Delta t$ must be less than $10^{-6}$; namely we require,

$$\frac{\sum_{x,y} |V_x(t + \Delta t) - V_x(t)|}{\sum_{x,y} |V_x^*|} \;<\; 10^{-6} \;\; \Delta t \tag{28}$$

and similarly for $V_y$.

In the case of transient flow such as the one-quarter Taylor vortex and the oscillating plate, we measure the error $V^E$ at the final time $T = 1.0$ using equations 20 and 21.

## 8.1  Boundary Implementation

The hybrid method d2q9H uses the standard collision operator at the inner nodes, and the extended collision operator at the boundary nodes. An important implementation issue is the calculation of the gradients of the fluid velocity at the boundary nodes. Below we will see that the best results are achieved when the gradients of the fluid velocity are specified as a Neumann boundary condition. This means that both the fluid velocity and the gradients of the fluid velocity are specified at the boundary nodes (both Dirichlet and Neumann boundary conditions). In practice however it is possible to specify only some of the velocity gradients at the boundary nodes, and not all of them. When a velocity gradient can not be specified, finite differences must be used to estimate it. For example the gradient $\partial V_x/\partial y$ at the top and bottom walls of the driven cavity problem [11, page 199] can not be specified as a Neumann condition because it is part of the solution that we seek to compute.

In our simulations we test the two extreme cases, the best case when Neumann boundary conditions are used to specify all of the velocity gradients, and the worst case when finite differences are used to calculate all of the velocity gradients at the boundary nodes. When Neumann conditions are used, we denote the lattice Boltzmann method by d2q9H$_{XD}$ ($XD$ stands for exact derivatives at the boundary). When first order asymmetric differences are used, we denote the method by d2q9H$_{1FD}$. When second order asymmetric differences are used, we denote the method by d2q9H$_{2FD}$. We will see that finite differences trigger instabilities when $M_c$ is large, and that first order differences are more stable than second order differences, but second order differences are more accurate when $M_c$ is small.

In our simulations we also test the lattice Boltzmann scheme d2q9F0 which uses the standard collision operator at every node, both boundary and inner nodes. At the boundary nodes the method d2q9F0 sets the populations $F_i$ equal to the equilibrium values $F_i^{eq}$ of the standard collision operator given by equation 6. At startup the method d2q9F0 initializes the $F_i$ equal to the equilibrium values $F_i^{eq}$ of the standard collision operator as described in section 7. In this section however, we use the extended collision operator for initialization in order to avoid large initial errors (see section 7), and we switch to the standard collision operator after the first step.

Regarding boundary conditions for the explicit finite difference projection method, we use Dirichlet conditions for the velocity, and Neumann conditions for the pressure $P$. In particular we require that $\partial P/\partial n = 0$ on the boundary, where $\partial n$ denotes the direction normal to the boundary [11, page 160]. The Neumann conditions are applied at the beginning of the SOR calculation, and the boundary values of the pressure $P$ are held constant throughout the SOR calculation.

Coming back to the lattice Boltzmann method, another issue regarding boundary conditions is the value of the density $\rho$ at the boundary. In many fluid flow cases it is appropriate to specify the value of $\rho$ at the boundary. For example this is done in the case of the one-quarter Taylor vortex and the Hagen-Poiseuille flow. In other fluid flow cases however it is appropriate to consider $\rho$ unknown at the boundary. In general there may be density gradients along the boundary that develop as a result of the fluid dynamics. For example in the driven cavity problem a density gradient develops along the walls of the cavity (pressure gradient divided by the square of the speed of sound) which is part of the fluid flow solution that we seek to compute. In such a case the density $\rho$ must be calculated **dynamically** from the simulated flow.

A good method of calculating the density $\rho$ at the boundary dynamically is to calculate $\rho$ as the average of the populations $F_i$ that "bring fluid into the boundary node" from neighboring nodes such as inner nodes and/or other neighboring boundary nodes. For example in the case of a horizontal wall that bounds the fluid region from below, the density $\rho$ must be calculated as the average of the populations $F_0, F_1, F_5, F_6, F_7, F_8$ when using an orthogonal grid (see appendix). The populations $F_2, F_3, F_4$ must be omitted in this calculation because they convect into the bottom wall from the outside of the fluid region. Similar calculations of the density must be done for all possible orientations of the boundary. In our simulations of the driven cavity problem and other flows past obstacles we have obtained good results (this is a qualitative judgement) using this approach.

## 8.2   Comparison — Boundary Conditions

In figure 9 we compare the methods d2q9H$_{XD}$, d2q9H$_{1FD}$, d2q9H$_{2FD}$, and d2q9F0 (solid, dashed, dotted, and dash-dotted lines) in simulations of the one-quarter Taylor vortex, the Hagen-Poiseuille flow, and the oscillating plate, figures (a), (b), (c) respectively. A $30 \times 30$ grid is used, and the error is plotted against $M_c$ with $\Delta t$ varying, and is calculated at the final time $T = 1.0$. The standard collision operator d2q9F0 achieves its smallest error when the relaxation parameter $\tau = 1$, at which point the standard and extended collision operators are identical. The hybrid method achieves its best results when Neumann conditions are used to calculate the velocity gradients at the boundary nodes (method d2q9H$_{XD}$). The use of finite differences at the boundary nodes (methods d2q9H$_{1FD}$ and d2q9H$_{2FD}$) leads to instabilities when $M_c$ is large. First order differences are more stable than second order differences, while second order differences are more accurate when $M_c$ is small.

## 8.3   Quadratic Convergence — Boundary Conditions

Figure 10 compares the lattice Boltzmann method d2q9H$_{XD}$ against the explicit finite difference projection method EP9 in simulations of the one-quarter Taylor vortex, the Hagen-Poiseuille flow, and the oscillating plate, figures (a), (b), (c) respectively. The error is plotted against the dimensionless parameter $\Delta t \, \nu / \Delta x^2$ to facilitate comparison between different grids. The curves correspond to d2q9H$_{XD}$ using $30 \times 30$ grid, d2q9H$_{XD}$ using $60 \times 60$ grid, EP9 using $30 \times 30$ grid, and EP9 using $60 \times 60$ grid (solid, dashed, dotted, dash-dotted lines). Figure (b) shows most clearly the rate of convergence in time. The lattice Boltzmann method has second order convergence in time (slope $-2$), and the finite difference projection method EP9 has first order convergence in time (slope $-1$). Both methods have second order convergence in space.

Furthermore we note that the use of first order differences to calculate the velocity gradients at the boundary nodes does not change the overall second order convergence of the lattice Boltzmann method. This can be seen in figure 10 (d) which corresponds to the same experiment as figure 10 (a) but uses the method d2q9H$_{1FD}$ (first order differences to calculate the velocity gradients at the boundary nodes) instead of the method d2q9H$_{XD}$ (Neumann conditions for the velocity at the boundary nodes). The second order convergence of the lattice Boltzmann method is discussed theoretically in references [8] and [9].

# Appendix: Orthogonal 9-speed Model

We consider an orthogonal lattice with nine populations at each node. The populations $F_i^{II}$ $i = 2, 4, 6, 8$ move along the diagonal directions at the speed $\sqrt{2}c$, while the populations $F_i^I$ $i = 1, 3, 5, 7$ move along the vertical and horizontal directions at the speed $c = \Delta x / \Delta t$. $F_0$ is the non-moving population. We denote the orthogonal 9-speed model with the symbol d2q9 following the convention of [2]. The relaxation and convection steps are given by the following formulas,

$$
\begin{aligned}
F_i(\vec{x} + \vec{e}_i \, \Delta t, \, t + \Delta t) &= F_i(\vec{x}, t) + (-1/\tau) \, [F_i(\vec{x}, t) - F_i^{\text{eq}}(\vec{x}, t)] \\
F_0(\vec{x}, \, t + \Delta t) &= F_0(\vec{x}, t) + (-1/\tau) \, [F_0(\vec{x}, t) - F_0^{\text{eq}}(\vec{x}, t)] \\
i &= 1, \ldots, 8
\end{aligned}
\tag{29}
$$

$$
\tau = \frac{1}{2} + \frac{3\Delta t \, \nu}{\Delta x^2} \quad .
$$

The relaxation parameter $\tau$ is chosen to achieve the desired kinematic viscosity $\nu$ given the space and time discretization parameters $\Delta x, \Delta t$. The vector $\vec{e}_i$ stands for the eight velocity directions of the orthogonal lattice,

$$
\vec{e}_i = \frac{\Delta x}{\Delta t} \left( \cos \frac{2\pi(i-1)}{8}, \, \sin \frac{2\pi(i-1)}{8} \right) \quad .
\tag{30}
$$

The velocity $\vec{V}(\vec{x}, t)$ and density $\rho(\vec{x}, t)$ are computed from the populations $F_i(\vec{x}, t)$ using the relations,

$$
\begin{aligned}
\rho(\vec{x}, t) &= \sum_{i=0}^{8} F_i(\vec{x}, t) \\
\rho(\vec{x}, t)\, \vec{V}(\vec{x}, t) &= \sum_{i=1}^{8} F_i(\vec{x}, t)\, \vec{e_i}
\end{aligned}
\tag{31}
$$

The variations of density around its mean value (spatial mean which is constant in time) provide an estimate of the fluid pressure $P(\vec{x}, t)$, according to the following equation,

$$
P(\vec{x}, t) = c_s^2\, (\rho(\vec{x}, t) - <\rho>) \quad .
\tag{32}
$$

The speed of sound is,

$$
c_s = \sqrt{(2\, w_0 + 4\, y_0)}\, (\Delta x/\Delta t)
\tag{33}
$$

where the coefficients $w_0, y_0$ are discussed below. The equilibrium populations $F_i^{\text{eq}}(x, t)$ are given by the following equations,

$$
\begin{aligned}
F_i^{\text{eq}II} &= \rho\, \left[ y_0 + y_1(\vec{e_i} \cdot \vec{V}) + y_{20}(\vec{e_i} \cdot \vec{V})(\vec{e_i} \cdot \vec{V}) + y_{21}(\vec{V} \cdot \vec{V}) \right] \\
F_i^{\text{eq}I} &= \rho\, \left[ w_0 + w_1(\vec{e_i} \cdot \vec{V}) + w_{20}(\vec{e_i} \cdot \vec{V})(\vec{e_i} \cdot \vec{V}) + w_{21}(\vec{V} \cdot \vec{V}) \right] \\
F_0^{\text{eq}} &= \rho\, \left[ z_0 + z_{21}(\vec{V} \cdot \vec{V}) \right]
\end{aligned}
\tag{34}
$$

$$
\begin{aligned}
&4\, w_0 + 4\, y_0 + z_0 = 1 \quad , \\
&y_1 = 1/(12\, c^2) \quad , \quad y_{20} = 1/(8\, c^4) \quad , \quad y_{21} = -1/(24\, c^2) \\
&w_1 = 1/(3\, c^2) \quad , \quad w_{20} = 1/(2\, c^4) \quad , \quad w_{21} = -1/(6\, c^2) \\
&z_{21} = -2/(3\, c^2) \quad , \quad c = \Delta x/\Delta t
\end{aligned}
$$

In our simulations we use $y_0 = (1/4)\, w_0$ and $w_0 = (1/7)$ unless otherwise indicated. The shear and bulk viscosity of the d2q9 collision operator have the following values (calculated using the Chapman-Enskog procedure),

$$
\nu = \frac{c^2\, \Delta t}{6}\, (2\tau - 1)
\tag{35}
$$

$$\mu = \frac{c^2 \Delta t}{3} (2\tau - 1)(1 - 3w_0 - 6y_0)$$

The extended collision operator (d2q9X) for the orthogonal 9-speed model is derived similarly to the hexagonal model of section 5. Two additional terms based on gradients of the fluid velocity are included in the equilibrium population formulas. Everything else, including all the coefficients $w_1, y_1, w_{20}, \ldots$ of the standard collision operator d2q9 remain the same. The equilibrium population formulas for d2q9X are as follows,

$$
\begin{aligned}
F_i^{\text{eq}II} &= \rho \left[ y_0 + y_1(\vec{e_i} \cdot \vec{V}) + y_{20}(\vec{e_i} \cdot \vec{V})(\vec{e_i} \cdot \vec{V}) + y_{21}(\vec{V} \cdot \vec{V}) \right] + \\
&\quad y_{31}(\vec{e_i} \cdot \vec{\nabla}(\vec{e_i} \cdot \rho \vec{V})) + y_{32}(\vec{\nabla} \cdot \rho \vec{V}) \\
F_i^{\text{eq}I} &= \rho \left[ w_0 + w_1(\vec{e_i} \cdot \vec{V}) + w_{20}(\vec{e_i} \cdot \vec{V})(\vec{e_i} \cdot \vec{V}) + w_{21}(\vec{V} \cdot \vec{V}) \right] + \\
&\quad w_{31}(\vec{e_i} \cdot \vec{\nabla}(\vec{e_i} \cdot \rho \vec{V})) + w_{32}(\vec{\nabla} \cdot \rho \vec{V}) \\
F_0^{\text{eq}} &= \rho \left[ z_0 + z_{21}(\vec{V} \cdot \vec{V}) \right] + z_{32}(\vec{\nabla} \cdot \rho \vec{V}) \\
2c^2 w_{31} + 4w_{32} &+ 4c^2 y_{31} + 4y_{32} + z_{32} = 0
\end{aligned}
\tag{36}
$$

The velocity gradients are computed using finite differences unless they are known by other means; for example the velocity gradients may be known at the boundary nodes (see section 8). In our simulations we use second order symmetric differences [11, page 19] at the inner nodes, and first or second order asymmetric differences at the boundary nodes as discussed in section 8. First order differences at the boundary are more stable for large $M_c$, while second order asymmetric differences at the boundary are more accurate for small $M_c$.

The shear and bulk viscosity of the d2q9X operator have the following values (calculated using the Chapman-Enskog procedure),

$$\nu = \frac{c^2 \Delta t}{6} (2\tau - 1) - c^4 w_{31} \tag{37}$$

$$\mu = \frac{c^2 \Delta t}{3} (2\tau - 1)(1 - 3w_0 - 6y_0) - 2c^4 w_{31} - 2c^2(w_{32} + 2y_{32})$$

In our simulations we use $y_{31} = w_{31}/4$ and $y_{32} = w_{32}/4$. Once the relaxation parameter $\tau$ is set equal to one, the coefficient $w_{31}$ is chosen to achieve the desired kinematic viscosity $\nu$ given the discretization parameters $\Delta x, \Delta t$. The coefficient $w_{32}$ is chosen to achieve the desired bulk viscosity. In the case of the hybrid method d2q9H, the bulk viscosity of equation 37 is chosen equal to the bulk viscosity of the standard collision operator given by equation 35. The coefficient $z_{32}$ must satisfy $2\,c^2\,w_{31} \; + \; 4\,w_{32} \; + \; 4\,c^2\,y_{31} \; + \; 4\,y_{32} \; + \; z_{32} \; = \; 0$ so that the equilibrium populations conserve mass.

Figure 1: The velocity field of the hexagonal Taylor vortex and the hexagonal shear flow are shown in figures (a) and (b) respectively. Both flows have periodic boundary conditions.

Figure 2: The four initialization methods d2q7F0, d2q7F1, d2q7X, d2q7H (solid, dashed, dotted, dash-dotted lines) are compared using a $30 \times 30$ grid and periodic boundary conditions. Figures (a) and (b) plot the error in simulating the hexagonal Taylor vortex using $\Delta t = 0.001$ and $\Delta t = 0.025$ respectively ($\tau = 0.5912$ and $\tau = 2.780$). Figures (c) and (d) plot the same data in the case of shear flow.

Figure 3: The extended collision operator is examined after many iterations. The error is plotted against $M_c$ with $\Delta t$ varying, and is calculated at the final time $T = 1.0$. The curves correspond to the hybrid method d2q7H, the extended collision operator d2q7X using finite differences to calculate the gradients, and again the extended collision operator d2q7X using the known analytic solution to calculate the gradients (solid, dashed, dotted lines). Figure (a) shows the error in simulating the hexagonal Taylor vortex, and figure (b) shows the error in simulating the hexagonal shear flow.

Figure 4: The error of the lattice Boltzmann method d2q7H is shown when single precision arithmetic is used, when single precision arithmetic together with the algebraic transformation of section 6 is used, and when double precision arithmetic is used (dotted, dashed, solid lines).

Figure 5: The error of the lattice Boltzmann method d2q7H is compared against the error of the explicit finite difference projection method EP7. The curves correspond to d2q7H using $30 \times 30$ grid, d2q7H using $60 \times 60$ grid, EP7 using $30 \times 30$ grid, and EP7 using $60 \times 60$ grid (solid, dashed, dotted, dash-dotted lines). Figures (a) and (b) show the error in simulating the hexagonal Taylor vortex, and figures (c) and (d) show the error in simulating the hexagonal shear flow.

Figure 6: The error of d2q7H is plotted against $M_c$ with $\Delta t$ varying, while keeping the Mach number $M$ constant and varying the density parameter $w_0$ (two dashed lines). For comparison purposes, the error of d2q7H when the Mach number varies and the density parameter $w_0 = 1/7$ is held constant is also shown (two solid lines). Results are shown for a $30 \times 30$ and a $60 \times 60$ grid. Figures (a), (b), (c) correspond to the hexagonal Taylor vortex at $M = 0.02$, the hexagonal Taylor vortex at $M = 0.1$, and the hexagonal shear flow at $M = 0.05$ respectively.

Figure 7: The error of d2q7H applied to the hexagonal Taylor vortex, and the error of d2q9H applied to the orthogonal Taylor vortex are shown (solid and dashed lines). In addition the error of the explicit finite difference projection method is shown when applied to the hexagonal Taylor vortex with $\Delta y = \Delta x \sqrt{3}/2$ and also the orthogonal Taylor vortex with $\Delta y = \Delta x$ (dotted and dash-dotted lines).

Figure 8: The velocity field of the one-quarter Taylor vortex, the Hagen-Poiseuille flow, and the oscillating plate problem are shown in figures (a), (b), (c) respectively. Boundary nodes are marked with a square. Figure (c) is plotted at time $t = 0.4$ when the oscillating plate starts moving to the left while the fluid below is still moving to the right.

Figure 9: The error of d2q9H$_{XD}$, d2q9H$_{1FD}$, d2q9H$_{2FD}$, and d2q9F0 (solid, dashed, dotted, and dash-dotted lines) is shown in simulations of the one-quarter Taylor vortex, the Hagen-Poiseuille flow, and the oscillating plate — figures (a), (b), (c) respectively.

Figure 10: The error of the lattice Boltzmann method d2q9H$_{XD}$ is compared against the error of the explicit finite difference projection method EP9. The curves correspond to d2q9H$_{XD}$ using $30 \times 30$ grid, d2q9H$_{XD}$ using $60 \times 60$ grid, EP9 using $30 \times 30$ grid, and EP9 using $60 \times 60$ grid (solid, dashed, dotted, dash-dotted lines). Figures (a), (b), (c) show simulations of the one-quarter Taylor vortex, the Hagen-Poiseuille flow, and the oscillating plate respectively. Figure (d) shows the same experiment as figure (a) using d2q9H$_{1FD}$ instead of d2q9H$_{XD}$.